# Spatial filtering of interlayer exciton ground state in WSe$_2$/MoS$_2$ heterobilayer

Disheng Chen[1,2,†], Kevin Dini[1,†], Abdullah Rasmita[1], Zumeng Huang[1], Qinghai Tan[1,2], Hongbing Cai[1,2], Ruihua He[3], Yansong Miao[3], Timothy C. H. Liew[1,4]*, Wei-bo Gao[1,2,4,5]*

[1]*Division of Physics and Applied Physics, School of Physical and Mathematical Sciences, Nanyang Technological University, Singapore 637371, Singapore.*

[2]*The Photonics Institute and Centre for Disruptive Photonic Technologies, Nanyang Technological University, Singapore 637371, Singapore.*

[3]*Institute For Digital Molecular Analytics and Science, Nanyang Technological University, Singapore 636921, Singapore.*

[4]*MajuLab, International Joint Research Unit UMI 3654, CNRS, Université Côte d'Azur, Sorbonne Université, National University of Singapore, Nanyang Technological University, Singapore 637371, Singapore.*

[5]*Centre for Quantum Technologies, National University of Singapore, Singapore 117543, Singapore.*

[†]*These authors contributed equally*

## Abstract

Long-life interlayer excitons (IXs) in transition metal dichalcogenide (TMD) heterostructure are promising for realizing excitonic condensates at high temperatures. Critical to this objective is to separate the IX ground state (the lowest energy of IX state) emission from other states' emissions. Filtering the IX ground state is also essential in uncovering the dynamics of correlated excitonic states, such as the excitonic Mott insulator. Here, we show that the IX ground state in WSe$_2$/MoS$_2$ heterobilayer can be separated from other states by its spatial profile. The emissions from different moiré IX modes are identified by their different energies and spatial distributions, which fits well with the rate-diffusion model for cascading emission. Our results show spatial filtering of the ground state mode and enrich the toolbox to realize correlated states at elevated temperatures.

**Main text**

The plethora of multiple phases in two-dimensional (2D) transition metal dichalcogenide (TMD) heterostructures with moiré superlattice makes it a promising platform to study many-body correlated states[1-5], including bosonic condensates[6] and excitonic Mott insulator[7,8]. In particular, the 2D heterobilayer with type-II band alignment, such as $WSe_2/MoS_2$ and $WSe_2/MoSe_2$, hosts interlayer excitons (IXs) with electron and hole separated in different layers[9,10], resulting in composite bosonic quasiparticles with a long lifetime reaching microseconds[9,11,12]. The long lifetime and the strong binding energy of these IXs make them suitable for realizing bosonic condensation at elevated temperatures[13]. Moreover, the moiré-induced localization results in moiré IX with enhanced on-site IX-IX repulsion, making this system a promising platform for simulating Bose-Hubbard model[14].

For the realization of IX condensate and the simulation of Bose-Hubbard model, a mechanism to isolate the IX ground state (i.e., the lowest energy of IX state) is crucial. Optical pumping of IXs involves exciting the intralayer excitons, which quickly decay into IXs due to the ultrafast interlayer charge transfer[15,16]. However, the existence of multiple moiré IX bands[17-24] allows the intralayer excitons to decay to higher-energy IX modes other than the IX ground state. It is thus crucial to have an efficient method to separate the ground state from other mode emissions.

In this work, we present evidence that the IX ground state emission in WSe$_2$/MoS$_2$ heterobilayer with strain-induced potential can be separated from other modes using spatial filtering. The IX PL spectrum features multiple peaks, indicating multiple IX bands, which correspond to multiple IX modes. We then conduct energy- and time-resolved IX spatial distribution measurements to study the IX state dynamics. Our experimental results show that each IX mode can be identified by its unique spatial distribution. Compared to previous experimental results on IX diffusion[25-28], our results show the first demonstration of spatial filtering of ground state moiré IX. Such findings are supported by the cascade transition rate-diffusion model, which fits the experimental data well.

Our sample consists of h-BN encapsulated WSe$_2$/MoS$_2$ heterobilayer. Monolayers WSe$_2$ and MoS$_2$, and thin hBN were mechanically exfoliated from their bulk crystal. The MoS$_2$/WSe$_2$ heterostructures were fabricated on the SiO$_2$/Si substrate with ultralow doping Si via a standard dry-transfer technique. The sharp edges of the exfoliated monolayers are aligned, resulting in commensurate stacking with 1° uncertainty in twist angle. Within this small twist angle range, the moiré superlattice constant is not sensitive to the twist angle (see Supplementary Note 1). These heterostructure samples were annealed under an ultrahigh vacuum (around $10^{-7}$ mbar) at 200 °C for 2 hours.

To obtain the IX PL emission, we excite our h-BN encapsulated WSe$_2$/MoS$_2$ heterobilayer sample using a 726 nm laser excitation, on-resonant with the WSe$_2$ intralayer trion energy[29] (see Methods for more details on optical measurement setup). Due to the ultrafast interlayer charge transfer[15,16,30], the WSe$_2$ intralayer trions quickly decay to become IXs[10,31,32], with holes in the WSe$_2$ layer and electrons in the MoS$_2$ layer (Fig. 1a, see also Supplementary

Information Fig. S4). These IXs can diffuse, resulting in PL emission diffusion profile, from which the IX spatial distribution can be obtained[28]. In our experiment, the IX spatial distribution is obtained by focusing a laser pulse excitation at one location and collecting the spatially resolved PL emission. Unless otherwise stated, our experimental results are obtained from the excitation location marked with a black circle in the PL intensity map shown in Fig. 1b. We found that the IX PL spectrum exhibits several PL peaks (Fig. 1c), indicating multiple IX energy bands, each with its own IX mode (labelled as mode 1, 2, and 3). We note that, while IX is sensitive to strain[33], it is unlikely that these peaks are due to disorder-induced strain (such as bubble) because PL from other locations also give similar peak energies (see Supplementary Information Fig. S5). Instead, they should be caused by the periodic modulation due to moiré superlattice potential (which can include strain modulation), i.e., they are different modes of moiré IX[19,34].

To characterize these IX modes and their dynamics, we performed time-resolved IX spatial distribution measurements while applying different wavelength filter configurations at the collection arm. The spectral-dependent IX spatial distribution is obtained by integrating the data over time (Fig. 1d, also see the energy- and time-resolved data in Supplementary Information Fig. S6), while the temporal evolution is studied by integrating the data over space (Fig. 1e). As shown in Fig. 1d, we observed that the high energy IX PL emission (i.e., mode 1) has a concentric profile, while the low energy IX PL emission (i.e., the sum of mode 2 and mode 3) has a ring-like profile. Additionally, PL spectra taken from different collection points (Supplementary Information Fig. S7) show that the emission from the centre (peripheral) locations is dominated by the high (low) energy IX mode, consistent with the previous statement. We note that the large energy difference (~ 140 meV) between the peak energy at these locations excludes the possibility of this difference being due to different density-induced repulsions (see Supplementary Note 2 for more discussion). The distinct profiles of each mode show that the IX spatial distribution can be used to identify the IX mode, and therefore we can filter out a specific IX mode by spatial filtering. A similar observation is obtained from different excitation locations in the sample (Supplementary Information Fig. S8) and from a different sample (Supplementary Information Fig. S9). In the time-resolved measurement, as shown in Fig. 1e, multiple exponential fittings are necessary to fit the PL intensity temporal evolution, which also corresponds to the three different IX modes. More data on time-resolved PL intensity under different filter configurations (Supplementary Information Fig. S10) also show that different modes have different lifetimes.

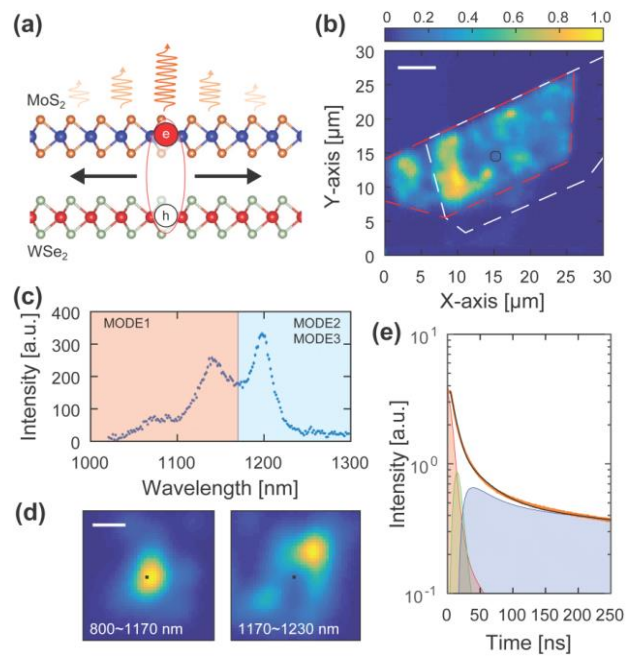

**Figure 1. Mode-dependent IX spatial distribution.** **(a)** $WSe_2/MoS_2$ interlayer exciton diffusion. The IX (electron in $MoS_2$ and hole in $WSe_2$) diffusion results in the temporal evolution of PL emission spatial distribution. **(b)** Interlayer exciton PL map of the stacked sample excited by 35 μW 726 nm CW laser while monitored through an 800 nm long-pass filter. The map is obtained by scanning the excitation location and collecting the PL from the excitation location. Red and white dashed lines indicate the boundaries of $WSe_2$ and $MoS_2$ monolayers. The black square circle indicates the excitation location used in subsequent experiments. The scale bar is 5 μm. **(c)** PL spectrum under 100 μW 726 nm CW excitation. The shaded regions indicate the spectral ranges in the two maps in (d). **(d)** Spectrally-filtered emission maps under 3.4 μW 726 nm pulse excitation. The 726 nm excitation is fixed at the centre of the map while scanning the detection spatially (as indicated by the white dot at the centre).

The spectral detection window is labelled on each map. The white scale bar is 2.5 μm. The data at each plot is normalized to the maximum PL count at that plot. **(e)** Time-resolved decay of the total PL. The orange dots are data. The black line is the fitting result using three modes, as indicated by the shaded areas under the curve (red: mode 1, green: mode 2, and blue: mode 3).

To quantitatively model the spatial distribution and temporal evolution of the IX modes, we consider combined rate-diffusion equations, including bosonic cascade transition[34,35] (see Methods for more details on rate-diffusion equation and fitting procedure). The resulting coupled differential equations can be expressed as

$$\begin{array}{cccc} & \text{(1) Diffusion} & \text{(2) Decay} & \text{(3) Cascade transition} \\ & \Downarrow & \Downarrow & \Downarrow \\ \dfrac{dn_i}{dt} = & \{D\Delta n_i + \nabla \cdot (n_i \nabla V)\} & - \Gamma_i n_i & - \left\{ \sum_{i<j} r_{ij} n_i (n_j + 1) - \sum_{i>j} r_{ji} n_j (n_i + 1) \right\}, \end{array} \quad (1)$$

where $n_i$ is the population of $i^{\text{th}}$ mode, $D$ is the diffusivity, $V = V(x, y)$ is the renormalized slowly-varying spatially dependent potential, $\Gamma_i$ is the spontaneous decay rate of the $i^{\text{th}}$ mode to vacuum, and $r_{ij}$ is the scattering rate from the $i^{\text{th}}$ to $j^{\text{th}}$ mode. Three different IX modes (i.e., $i, j \in \{1, 2, 3\}$) are considered, with lower-numbered modes having higher energy. We note here that the potential $V$ does not include the moiré superlattice potential. Instead, it is due to strain and disorder in the sample. On the other hand, the moiré superlattice potential results in multiple moiré IX states and a $D$ value of around 0.1 cm²/s[25-28].

Based on this model, the mode-dependent IX spatial distribution and temporal evolution can be described as follows (see Fig. 2). The optical pump excites the free (i.e., unlocalized) intralayer excitons, which quickly decay to IXs due to ultrafast charge transfer. The less-localized nature of higher-energy IX modes makes it more likely that the unlocalized intralayer excitons decay to these modes. Hence, the optical excitation mainly results in the population of mode 1, which is the highest-energy IX mode with a short lifetime. The IXs then undergo three different processes, as expressed in Eq. (1): (1) diffusion affected by potential $V$, (2) spontaneous decay resulting in the PL signal, and (3) cascade transition to lower-energy modes (mode 2 and mode 3). As a result, the spatial profile starts to be affected by the shape of potential $V$ while the IX population starts to be dominated by mode 2, and, eventually, mode 3 with a long lifetime.

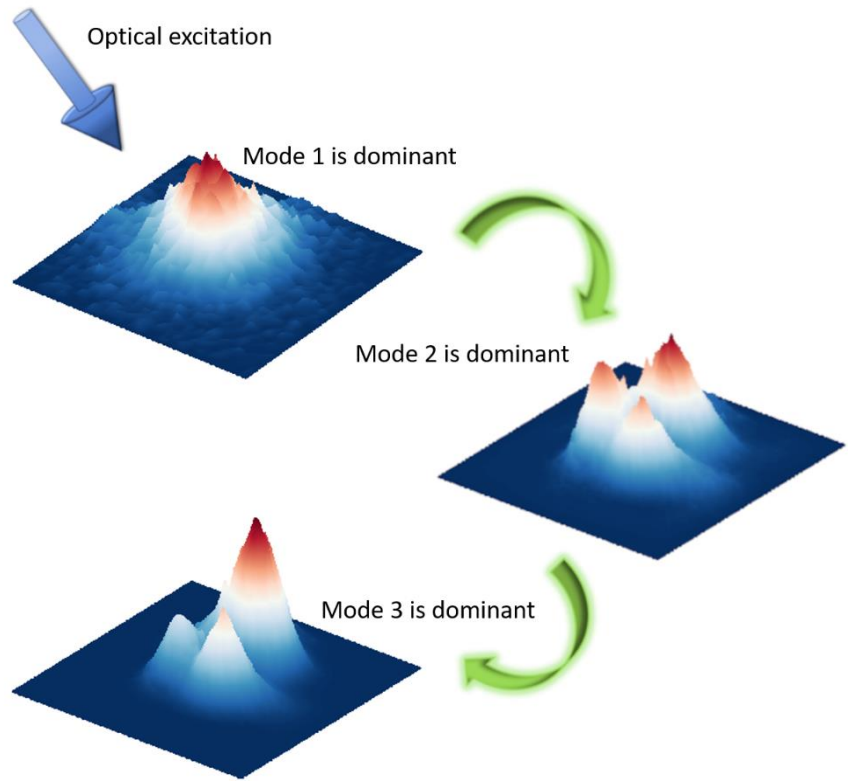

**Figure 2. Evolution of IX spatial distribution.** Optical excitation mainly results in population of IX mode 1, which then cascades to lower-energy modes (mode 2 and mode 3). The PL emission is dominated by high-energy mode 1 at the short time scale, while it is dominated by lower energy modes at the longer delay time. At the long time scale, the PL emission only consists of the lowest energy mode 3. The content of this figure is obtained from the simulated spatial distribution of each mode. The color scale in this figure corresponds to the normalized amplitude, with red is 1 and black is 0.

Figure 3 shows the time-resolved spatial distribution. The first and second rows show the experimental data and the simulation result, respectively. Each experimental frame is obtained by sweeping the collection arm across the sample while keeping the excitation spot fixed at the centre of the map. The excitation is implemented by a 70-ps pulsed laser centred at 726 nm wavelength with a repetition rate of 1 MHz. The fluorescence is photon counted with 64-ps time resolution for an integration period of 8 s using a time-correlated single photon counting (TCSPC) device, which has been synchronized to the excitation laser. The modes contributions obtained from the simulation are shown in the third row. As shown in the first and second rows of Fig. 3, the model above fits the time-resolved IX PL intensity and spatial distribution very well. As shown in the third row, the model shows that the higher-energy mode has a concentric spatial distribution, while the low-energy mode has a ring-like spatial distribution following the potential shape, agreeable with our data (Fig. 1d and Supplementary Information Fig. S6). At a short time scale (less than 1 ns), the emission is dominated by high-energy IX (mode 1, indicated by red colour). With increasing delay time, the emission from the lower-energy IXs mode 2 (green colour) and mode 3 (blue colour) dominates over mode 1. Eventually, at a long time scale (> 40 ns), the emission only consists of mode 3 with a long lifetime and a spatial profile reflecting the shape of the spatial dependent potential (Supplementary Information Fig. S11). More analysis showing the IX diffusion nature is presented in Supplementary Note 3.

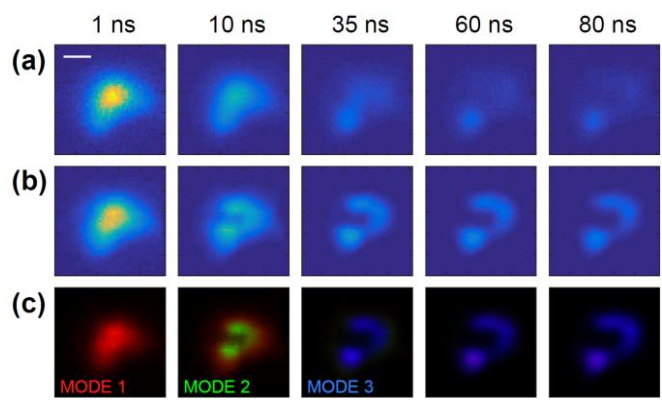

**Figure 3. Time-resolved spatial distribution of IX modes and their simulation.** (a) Experimental result. The data is normalized to the maximum of the entire evolution. A 726 nm pulsed excitation (1 MHz repetition) at 3.4 μW average power is used. Each frame is the spatial distribution of the PL signal with an exposure time of 64 ps. The scale bar is 2.5 μm. (b) Simulation result. The simulated spatial distribution is obtained by summing up the contribution from all modes together before normalizing it to the global maximum of the evolution. (c) Simulated

mode contribution. The relative magnitude of the three modes, which are colour coded as red, green, and blue, for modes 1, 2, and 3, respectively, normalized to the maximum intensity of each frame. More data at different time is presented in Supplementary Information Fig. S12.

We further characterize the power dependence of the IX spatial distribution. The experimental conditions and protocols are the same as those employed in Fig. 3(a), except that all frames within 1 μs period are summed up to give an averaged emission map as shown in Fig. 4 top row. We can see that the IX spatial distribution resembles the high (low)-energy mode 1 (3) at low (high) excitation power. The simulation results (Fig. 4 bottom row) also show similar behaviour. This behaviour can be attributed to the contribution of nonlinear processes. In particular, as the power increases, the contribution of the nonlinear cascade terms (i.e., the $n_i n_j$ terms in Eq. (1)) becomes more dominant. Consequently, the cascade decay rate to the low-energy modes is increased, resulting in the low-energy mode dominating the PL at high excitation power. More data on power dependence and temporal evolution at different temperatures support our findings (see Supplementary Information Fig. S13 and S14).

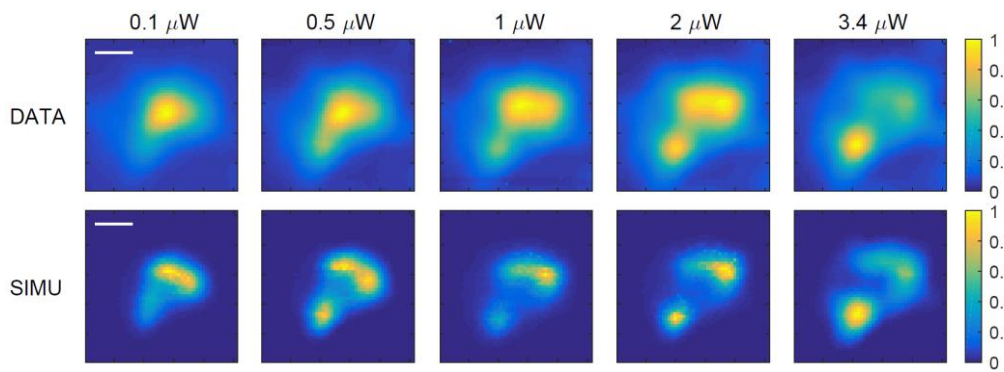

**Figure 4. Power-dependent spatial distribution of the PL emission.** Each map is taken by sweeping the collection path across the sample while keeping the excitation spot (of <1 um diameter) fixed at the centre of the map. The sample is excited by a 726 nm pulsed laser (1 MHz repetition rate) with an average power labelled on the top. The scale bar is 2.5 um. The data at each plot is normalized to the maximum PL count at that plot.

**Conclusion**

In conclusion, we have demonstrated that the ground state IX can be spatially distinguished from higher-energy IX modes. Such mode spatial filtering can be crucial for realizing effective ground state pumping towards achieving IX condensate at elevated temperatures. The moiré superlattice and externally induced strain can influence the spatially dependent potential. Further development of our results, including these effects, can open a pathway in engineering bosonic correlated states in the 2D heterostructure.

**Methods**

Optical measurement setup

All optical measurements were implemented at the liquid helium temperature of 4 K by placing the sample in a millikelvin (mK) dilution refrigerator. Two galvo scanners have been set up in the optical setup. Galvo 1, placed in the shared path of the excitation beam and the collection beam, provides joint controls over the directions of both beams simultaneously. The scan with Galvo 1 produces a PL map like Fig. 1b, where the collected fluorescence is always from the same location as the excitation spot. In comparison, Galvo 2 is placed only in the collection arm, thus only manipulating the collection path, which allows for studying the emission pattern from a specific excitation spot. To increase the scanning range of the galvos, we constructed a 4-f system with two achromatic lenses of f = 200 mm and f = 75 mm, respectively, which provides an extra 2.7x magnification in imaging. In cryostat, the PL from the sample was firstly collected and collimated by an objective with NA 0.9, and then filtered by an 808nm long-pass filter before being coupled into a single-mode fibre which directs the photons to a superconducting single-photon detector for photon counting measurements or to a spectrometer for spectrum analysis. The photon counting was accomplished by either an NI DAQ board or a time-correlated single photon counting (TCSPC) device (Picoquant Hydraharp). The latter has been employed in time-resolved measurements by synchronizing to a picosecond pulsed laser (Picoquant LDH-D-C-730). The collected data were finally processed with home-written code to obtain time-resolved maps, as shown in Fig. 3(a), or power-dependent emission patterns, as shown in Fig. 4(a).

Rate-diffusion model and fitting procedure

To describe the evolution of the exciton population in the three different coupled modes, we here put ourselves in the framework of incoherent populations and therefore work with rate equations. The spatial dynamics of the system is considered through the simple one-mode Smoluchowski diffusion equation:

$$\frac{dn_i}{dt} = D\Delta n_i + \nabla \cdot (n_i \nabla V), \qquad (2)$$

where $n_i$ is the population of $i^{\text{th}}$ mode, $D$ is the diffusivity, and $V = V(x,y)$ is the renormalized spatially dependent potential. We assume that the diffusion coefficient and the renormalized potential are the same for all three modes.

We now need to consider the two bosonic cascade mechanisms that coupled the three modes. The first one is spontaneous relaxation which is linear with the population of excitons. The second one is exciton-exciton scattering, which is nonlinear and depends on both the population of the initial and final states. Considering these mechanisms, the rate-diffusion equations can be expressed as in Eq. (1) in the main text.

To reduce the number of fitting parameters and increase the fit's precision, we need to obtain the shape of the spatially dependent potential. To do so, we assume that at a long time scale, i.e., after hundreds of ns, only the lowest energy mode is macroscopically populated, i.e., $n_1 = n_2 = 0$. Also, we can assume that the system has reached its spatial steady state and only the overall amplitude of the distribution changes. Therefore, the rate equation for $n_3$ can be expressed as:

$$\frac{dn_3}{dt} = -\Gamma_3 n_3, \quad (3)$$

leading to

$$D\Delta n_3 + \nabla \cdot (n_3 \nabla V) = 0. \quad (4)$$

In Eq. (4), $n_3 = n_3(x,y)$ is obtained from the measurement of the PL intensity distribution at the long time scale. We then used a regression algorithm to solve Eq. (4) for $V(x,y)$ with $D$ as a free parameter. The algorithm converges to a solution only if the given initial values are not too far from the solution. Therefore we used an initial guess of $V(x,y)$ proportional to $1/n_3(x,y)$. We averaged the algorithm over 1000 iterations with small random fluctuations.

Using the $V(x,y)$ and $D$, we calculated the scattering rates between the different modes and the decay rates. To do so, we first spatially integrate the signal to obtain the time dependence of the total exciton population. Then we fit the data using a regression algorithm considering $\left(n_1^0, n_2^0, n_3^0, \Gamma_1, \Gamma_2, \Gamma_3, r_{12}, r_{13}, r_{23}\right)$ as the variables where $n_i^0$ is the initial population of the $i^{th}$ mode after the pulse excitation (i.e., at $t = 0\,\text{s}$), $\Gamma_i$ is the spontaneous decay rate of the $i^{th}$ mode to vacuum, and $r_{ij}$ is the scattering rate from the $i^{th}$ to $j^{th}$ mode. We assume that the shape of the total intensity at $t = 0\,\text{s}$ measured experimentally is close to the spatial distribution of the first mode. We average over $10^4$ iterations with random fluctuations of up to 50% around the initial conditions. The values of the diffusion coefficient and the rates are shown in Supplementary Information Table S1.

We note that the discrepancy between the experimental result and simulation (main text Fig 3a and 3b), especially at the short time scale, is mainly due to the diffusion coefficient chosen to be the same for all modes in the simulation. In reality, the diffusion coefficient for different moiré IX modes can be different due to different effective masses. This approximation that all modes have the same effective mass was done to limit the degrees of freedom for the optimization algorithm and to keep the model simple.

## Data availability

The data that support the findings of this study are available from the corresponding authors upon reasonable request.

## Acknowledgements

This work is supported by the Singapore National Research Foundation through its Competitive Research Program (CRP Award No. NRF-CRP22-2019-0004, NRF-CRP23-2019-0002), QEP program and Singapore Ministry of Education (MOE2016-T3-1-006 (S)). K. D. and T. C. H. L. were supported by the Singapore Ministry of Education via Tier 2 Grant (MOE-T2EP50121-0020) and Tier 3 Grant (MOE2018-T3-1-002).

## Author contributions

D.C. performed the PL measurement with the help of Z.H. and Q.T., H.C. and Q.T. fabricated the devices, D.C. and K.D. analyzed the data, K.D. performed the theoretical analysis, K.D., A.R., and D.C. wrote the manuscript with inputs from all authors. T.C.H.L. and W-b.G. supervised the project. All authors contributed to the discussion of the results.

## Competing interests

The authors declare no competing interests.

## Additional information

**References**

1   Xu, Y. *et al.* Correlated insulating states at fractional fillings of moiré superlattices. *Nature* **587**, 214-218, doi:10.1038/s41586-020-2868-6 (2020).

2   Wang, L. *et al.* Correlated electronic phases in twisted bilayer transition metal dichalcogenides. *Nat. Mater.* **19**, 861-866, doi:10.1038/s41563-020-0708-6 (2020).

3   Tang, Y. *et al.* Simulation of Hubbard model physics in $WSe_2/WS_2$ moiré superlattices. *Nature* **579**, 353-358, doi:10.1038/s41586-020-2085-3 (2020).

4   Shimazaki, Y. *et al.* Strongly correlated electrons and hybrid excitons in a moiré heterostructure. *Nature* **580**, 472-477, doi:10.1038/s41586-020-2191-2 (2020).

5   Regan, E. C. *et al.* Mott and generalized Wigner crystal states in $WSe_2/WS_2$ moiré superlattices. *Nature* **579**, 359-363, doi:10.1038/s41586-020-2092-4 (2020).

6   Wang, Z. *et al.* Evidence of high-temperature exciton condensation in two-dimensional atomic double layers. *Nature* **574**, 76-80, doi:10.1038/s41586-019-1591-7 (2019).

7   Xiong, R. *et al.* Correlated insulator of excitons in $WSe_2/WS_2$ moiré superlattices. *Science* **380**, 860-864, doi:doi:10.1126/science.add5574 (2023).

8   Lagoin, C., Suffit, S., Baldwin, K., Pfeiffer, L. & Dubin, F. Mott insulator of strongly interacting two-dimensional semiconductor excitons. *Nat. Phys.* **18**, 149-153, doi:10.1038/s41567-021-01440-8 (2022).

9   Rivera, P. *et al.* Observation of long-lived interlayer excitons in monolayer $MoSe_2$–$WSe_2$ heterostructures. *Nat. Commun.* **6**, 6242, doi:10.1038/ncomms7242 (2015).

10  Karni, O. *et al.* Infrared interlayer exciton emission in $MoS_2/WSe_2$ heterostructures. *Phys. Rev. Lett.* **123**, 247402, doi:10.1103/PhysRevLett.123.247402 (2019).

11   Jiang, C. *et al.* Microsecond dark-exciton valley polarization memory in two-dimensional heterostructures. *Nat. Commun.* **9**, 753, doi:10.1038/s41467-018-03174-3 (2018).

12   Kim, J. *et al.* Observation of ultralong valley lifetime in $WSe_2/MoS_2$ heterostructures. *Sci. Adv.* **3**, e1700518, doi:doi:10.1126/sciadv.1700518 (2017).

13   Fogler, M. M., Butov, L. V. & Novoselov, K. S. High-temperature superfluidity with indirect excitons in van der Waals heterostructures. *Nat. Commun.* **5**, 4555, doi:10.1038/ncomms5555 (2014).

14   Park, H. *et al.* Dipole ladders with large Hubbard interaction in a moiré exciton lattice. *Nat. Phys.*, doi:10.1038/s41567-023-02077-5 (2023).

15   Jin, C. *et al.* Ultrafast dynamics in van der Waals heterostructures. *Nat. Nanotechnol.* **13**, 994-1003, doi:10.1038/s41565-018-0298-5 (2018).

16   Zimmermann, J. E. *et al.* Ultrafast charge-transfer dynamics in twisted $MoS_2/WSe_2$ heterostructures. *ACS Nano* **15**, 14725-14731, doi:10.1021/acsnano.1c04549 (2021).

17   Yu, H., Liu, G.-B., Tang, J., Xu, X. & Yao, W. Moiré excitons: From programmable quantum emitter arrays to spin-orbit-coupled artificial lattices. *Sci. Adv.* **3**, e1701696, doi:doi:10.1126/sciadv.1701696 (2017).

18   Brem, S., Linderälv, C., Erhart, P. & Malic, E. Tunable phases of moiré excitons in van der Waals heterostructures. *Nano Lett.* **20**, 8534-8540, doi:10.1021/acs.nanolett.0c03019 (2020).

19   Tran, K. *et al.* Evidence for moiré excitons in van der Waals heterostructures. *Nature* **567**, 71-75, doi:10.1038/s41586-019-0975-z (2019).

20   Seyler, K. L. *et al.* Signatures of moiré-trapped valley excitons in $MoSe_2/WSe_2$ heterobilayers. *Nature* **567**, 66-70, doi:10.1038/s41586-019-0957-1 (2019).

21	Jin, C. *et al.* Observation of moiré excitons in WSe$_2$/WS$_2$ heterostructure superlattices. *Nature* **567**, 76-80, doi:10.1038/s41586-019-0976-y (2019).

22	Alexeev, E. M. *et al.* Resonantly hybridized excitons in moiré superlattices in van der Waals heterostructures. *Nature* **567**, 81-86, doi:10.1038/s41586-019-0986-9 (2019).

23	Wu, F., Lovorn, T. & MacDonald, A. H. Theory of optical absorption by interlayer excitons in transition metal dichalcogenide heterobilayers. *Phys. Rev. B* **97**, 035306, doi:10.1103/PhysRevB.97.035306 (2018).

24	Choi, J. *et al.* Twist angle-dependent interlayer exciton lifetimes in van der Waals heterostructures. *Phys. Rev. Lett.* **126**, 047401, doi:10.1103/PhysRevLett.126.047401 (2021).

25	Choi, J. *et al.* Moiré potential impedes interlayer exciton diffusion in van der Waals heterostructures. *Sci. Adv.* **6**, eaba8866, doi:doi:10.1126/sciadv.aba8866 (2020).

26	Yuan, L. *et al.* Twist-angle-dependent interlayer exciton diffusion in WS$_2$–WSe$_2$ heterobilayers. *Nat. Mater.* **19**, 617-623, doi:10.1038/s41563-020-0670-3 (2020).

27	Jauregui, L. A. *et al.* Electrical control of interlayer exciton dynamics in atomically thin heterostructures. *Science* **366**, 870-875, doi:doi:10.1126/science.aaw4194 (2019).

28	Wang, J. *et al.* Diffusivity reveals three distinct phases of interlayer excitons in MoSe$_2$/WSe$_2$ heterobilayers. *Phys. Rev. Lett.* **126**, 106804, doi:10.1103/PhysRevLett.126.106804 (2021).

29	Li, Z. *et al.* Revealing the biexciton and trion-exciton complexes in BN encapsulated WSe$_2$. *Nat. Commun.* **9**, 3719, doi:10.1038/s41467-018-05863-5 (2018).

30	Zhu, H. *et al.* Interfacial charge transfer circumventing momentum mismatch at two-dimensional van der Waals heterojunctions. *Nano Lett.* **17**, 3591-3598, doi:10.1021/acs.nanolett.7b00748 (2017).

31  Tan, Q. *et al.* Layer-engineered interlayer excitons. *Sci. Adv.* **7**, eabh0863, doi:doi:10.1126/sciadv.abh0863 (2021).

32  Fang, H. *et al.* Strong interlayer coupling in van der Waals heterostructures built from single-layer chalcogenides. *Proc. Natl. Acad. Sci. USA* **111**, 6198-6202, doi:doi:10.1073/pnas.1405435111 (2014).

33  Cho, C. *et al.* Highly strain-tunable interlayer excitons in $MoS_2$/$WSe_2$ heterobilayers. *Nano Lett.* **21**, 3956-3964, doi:10.1021/acs.nanolett.1c00724 (2021).

34  Tan, Q., Rasmita, A., Zhang, Z., Novoselov, K. S. & Gao, W.-b. Signature of cascade transitions between interlayer excitons in a moiré superlattice. *Phys. Rev. Lett.* **129**, 247401, doi:10.1103/PhysRevLett.129.247401 (2022).

35  Liew, T. C. H. *et al.* Proposal for a bosonic cascade laser. *Phys. Rev. Lett.* **110**, 047402, doi:10.1103/PhysRevLett.110.047402 (2013).

**Supporting Information**

Discussion on twist angle dependence, IX-IX repulsion, and IX diffusion; Illustration of IX generation mechanism; PL spectrum dependence on excitation power, excitation location, and collection location; Time-resolved PL intensity dependence on collection location and IX mode; Mode dependent IX diffusion profile from other excitation locations and from another sample; Time-resolved spatial distribution dependence on IX mode and temperature; Power dependent spatial distribution at 50 mK; Renormalized potential profile.

# Supplementary information

# Spatial filtering of interlayer exciton ground state in WSe$_2$/MoS$_2$ heterobilayer

Disheng Chen[1,2,†], Kevin Dini[1,†], Abdullah Rasmita[1], Zumeng Huang[1], Qinghai Tan[1,2], Hongbing Cai[1,2], Ruihua He[3], Yansong Miao[3], Timothy C. H. Liew[1,4]*, Wei-bo Gao[1,2,4,5]*

[1]*Division of Physics and Applied Physics, School of Physical and Mathematical Sciences, Nanyang Technological University, Singapore 637371, Singapore.*

[2]*The Photonics Institute and Centre for Disruptive Photonic Technologies, Nanyang Technological University, 637371 Singapore, Singapore.*

[3]*Institute For Digital Molecular Analytics and Science, Nanyang Technological University, Singapore, Singapore.*

[4]*MajuLab, International Joint Research Unit UMI 3654, CNRS, Université Côte d'Azur, Sorbonne Université, National University of Singapore, Nanyang Technological University, Singapore, Singapore.*

[5]*Centre for Quantum Technologies, National University of Singapore, Singapore 117543, Singapore.*

[†]*These authors contributed equally*

# Table of contents

## Supplementary Note 1. Twist angle dependence of WSe$_2$/MoS$_2$ moiré superlattice

The moiré superlattice constant of the WSe$_2$/MoS$_2$ heterobilayer is given by $a_{Moire}(\theta) = a_{Se}/\sqrt{\delta^2 + \theta^2}$, where $\delta = |a_{Se} - a_S|/a_S$, $a_{Se} = 3.28$ Å and $a_S = 3.16$ Å are WSe$_2$ and MoS$_2$ lattice constants, respectively, and $\theta$ is the twist angle in radians[1]. The moiré superlattice density is given by $n_0 = 2/(a_{Moire}^2 \sqrt{3})$. The twist angle dependence is plotted in Fig. S1. As can be seen from this description, the lattice mismatch between WSe$_2$ and MoS$_2$ makes the angle dependence less important, especially for the small twist angle ($\leq 1°$ from the commensurate AA or AB stacking) case.

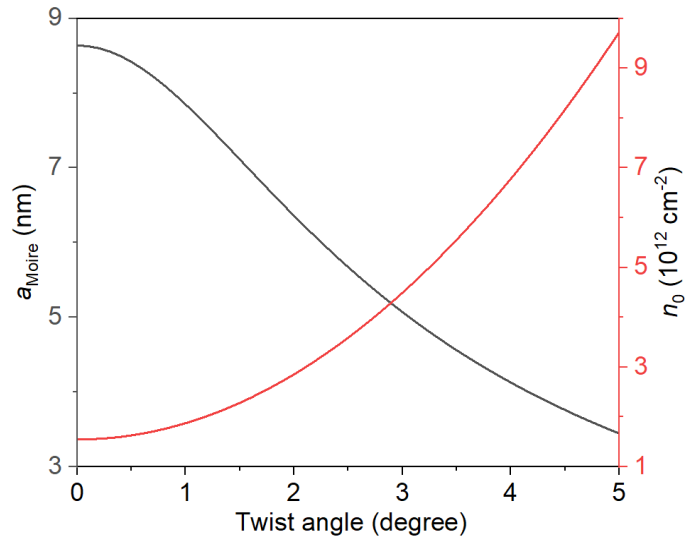

**Figure S1. Moiré superlattice vs twist angle.** The calculated moiré superlattice constant (black line) and density (red line) as a function of twist angle are shown.

## Supplementary Note 2. Discussion on IX-IX repulsion contribution

We note that a ring-like diffusion profile has been observed in intralayer exciton emission in monolayer WS$_2$ and associated with the Auger recombination contribution[2]. However, in this section, we show that the same mechanism cannot explain the ring-like profile of the low energy IX PL emission (Fig. 1d in the main text) observed in our results.

The main mechanism behind the Auger recombination contribution is that, at the excitation location, the large exciton density will increase the nonradiative decay rate, reducing the PL yield. As a result, for a high enough excitation power, the maximum intensity may happen at some distance from the excitation location, creating a ring-like emission profile. However, in this case, the peak energy difference between the centre and peripheral location will be limited by the IX-IX repulsion, which is less than 40 meV[3,4]. Our power-dependent PL spectrum measurement (Fig. S2) further agrees with this value. On the other hand, the observed energy difference in our results is ~ 140 meV, much larger than the IX-IX repulsion. Hence, the explanation based on Auger recombination cannot capture the main mechanism behind the observed phenomenon. Instead, as explained in the main text, this phenomenon can be attributed to the cascade transition between different IX modes.

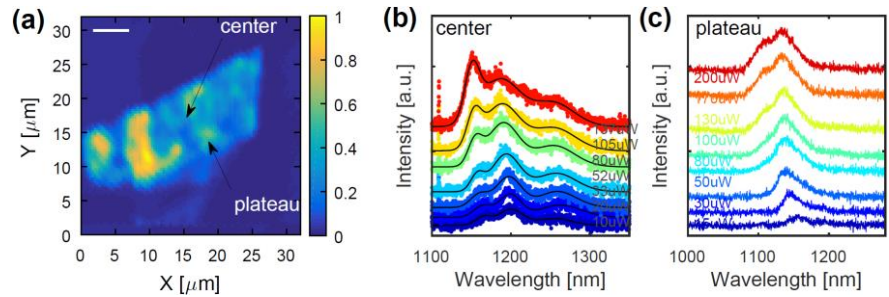

**Figure S2. Power dependence of the PL spectrum. (a)** PL intensity map showing the excitation locations used in b (center) and c (plateau). **(b, c)** Power dependent spectra. The observed blueshift between the spectrum at high excitation power is less than 7 meV.

### Supplementary Note 3. Analysis of IX diffusion

Figure S3b shows the time-dependent PL intensity at several collection points (marked as 1 to 4 in Fig. S3a), with the excitation location fixed at location 1 in Fig. S3a. We note that the excitation spot size is ~ 1 μm, much smaller than the distance between these four points, so locations 2 to 4 can be treated as far away from the excitation location. Initially, the count at the excitation location was higher than that at other locations. However, after some time, the count at other locations becomes higher than at the excitation location. Such behaviour indicates the diffusion of the interlayer excitons.

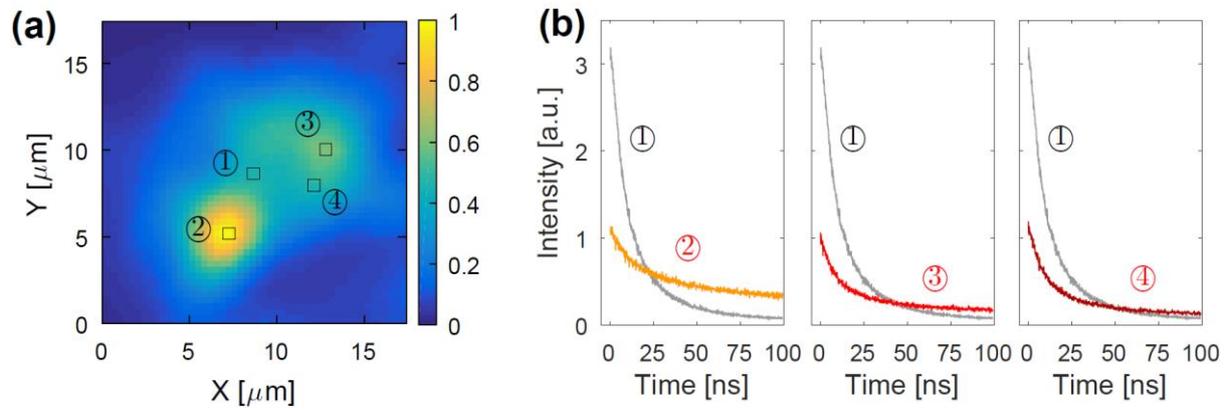

**Figure S3. Evidence of IX diffusion from temporal evolution from various collection points.** (a) The PL emission distribution when fixing the excitation spot at the center of the map. The excitation is implemented with a 3.4 µW 726 nm ps pulsed laser with 1 MHz repetition rate. (b) Comparison between the time-resolved PL collected at excitation location and far away from excitation location. The numbers 1 to 4 indicate the collection location labelled in a. After some time, the counts at the faraway locations become higher than at the excitation location.

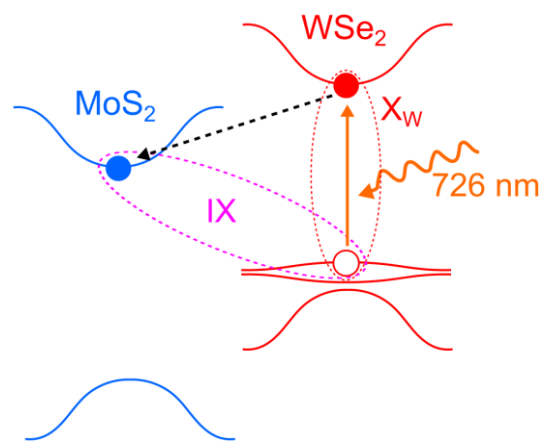

**Figure S4. Formation of IX in WSe$_2$/MoS$_2$ heterostructure under 726 nm excitation.** The 726 nm laser excites WSe$_2$ intralayer exciton ($X_W$), which then relaxes to interlayer exciton (IX) following the charge transfer process.

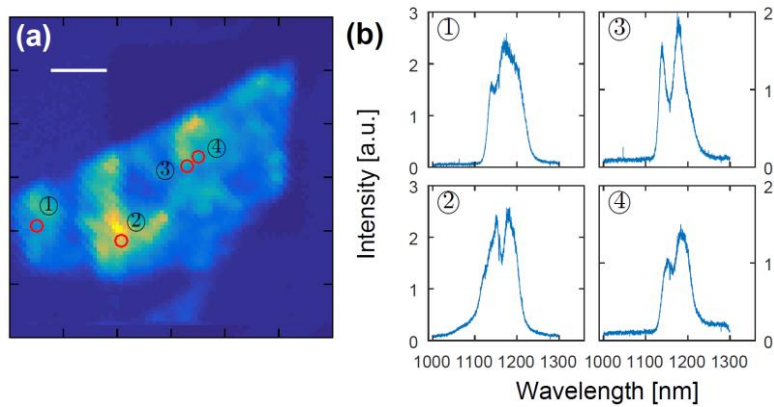

**Figure S5. PL spectrum from different excitation locations.** The scale bar in (a) is 5 μm. The spectra are taken under 100 μW CW excitation. All four spectra showed peaks at similar wavelengths.

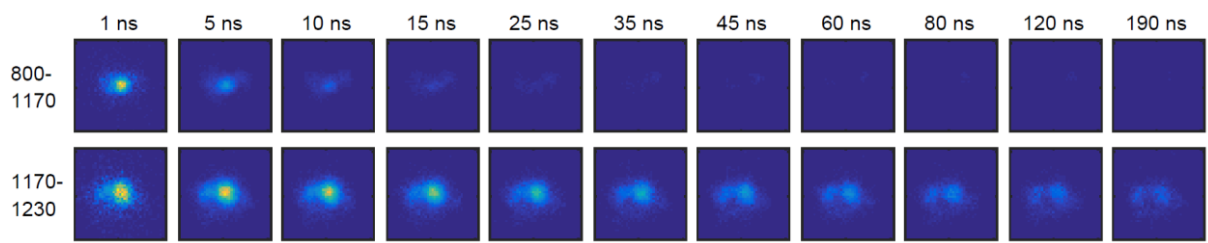

**Figure S6. Time-resolved spatial distribution of IX modes.** The data for high-energy emission (wavelength: 800 – 1170 nm, top row) and low-energy emission (1170 – 1230 nm, bottom row) are shown. The data is normalized to the maximum of the entire evolution. A 726 nm pulsed excitation (1 MHz repetition) at 3.4 μW average power is used.

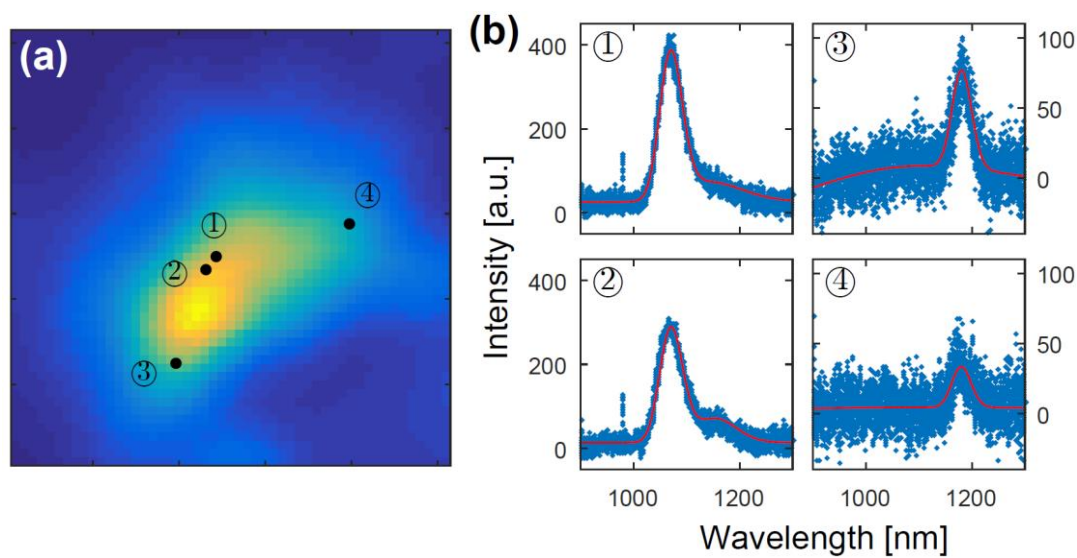

**Figure S7. PL spectrum at several collection locations.** The excitation location is fixed at the centre of the map in (a). The emission from the centre (peripheral) locations is dominated by the high (low) energy IX mode. The map (a) was taken with 726 nm pulsed excitation (1 MHz repetition) at 2.0 μW average power. The spectra 1 to 4 were taken with CW 726 nm laser at 130 uW for an exposure time of 10 seconds. An 808 nm long pass filter was implemented during data collection.

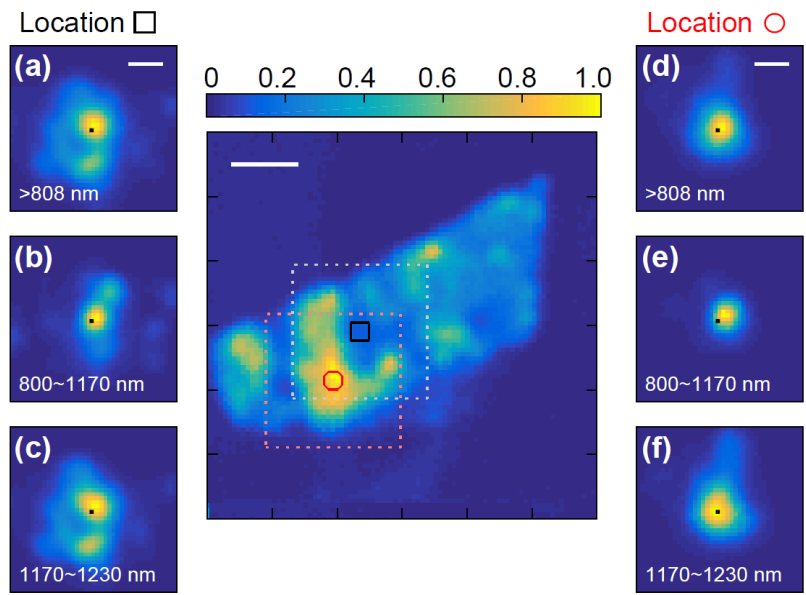

**Figure S8. Mode-dependent IX diffusion profile obtained from other excitation locations.** Middle: Normalized interlayer exciton PL map as in Fig. 1b. The excitation locations are indicated by the black box for (a-c) and the red hexagon (d-f). The scale bar in the middle map is 5 μm. **(a-c)** Spatial distributions of the PL when fixing the excitation at the centre of these maps while detecting through various filters. The maps are normalized to the maximum intensity of (a), which is the sum of all fluorescence. **(d-f)** Like (a-c) but from a different excitation location. The maps are normalized to the maximum intensity of (d), which is the sum of all fluorescence. The scale bars in (a-f) are 2.5 μm. Black spot in each map in (a-f) corresponds to the excitation spot. The grey dotted box in the middle figure gives the map range of (a-c), while the orange dotted box gives the map range of (d-f).

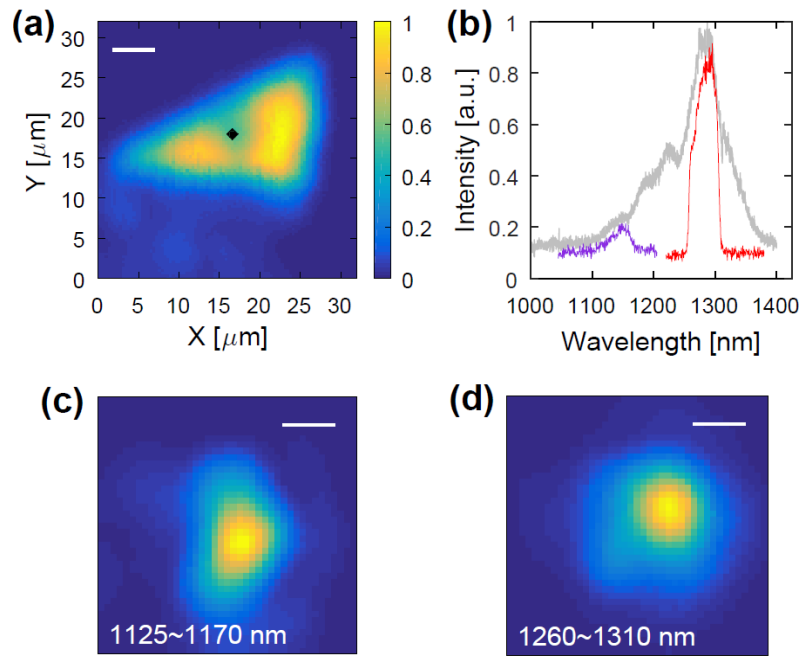

**Figure S9. Mode-dependent IX diffusion profile obtained from another sample. (a)** Interlayer exciton PL map of the stacked sample excited by 35 µW 726 nm CW laser while monitored through an 800 nm long-pass filter. The black diamond indicates the excitation location used in (b-d). The scale bar is 5 µm. **(b)** PL spectrum under 3.4 µW 726 nm pulse excitation. The spectra of the filtered emission used in (c, d) are shown together with the unfiltered spectrum. **(c, d)** Spectrally-filtered emission maps. The excitation is fixed at the centre of the map while scanning the detection spatially (as indicated by the white dot at the centre). The spectral detection window is labelled on each map. The white scale bar is 2.5 µm. The data at each plot is normalized to the maximum PL count at that plot.

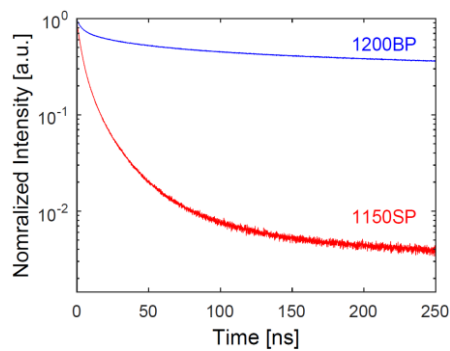

**Figure S10. Mode-dependent time-resolved PL.** The low-energy modes (blue) have a longer lifetime than the high-energy mode (red), agreeable with the simulation in Fig. 1e.

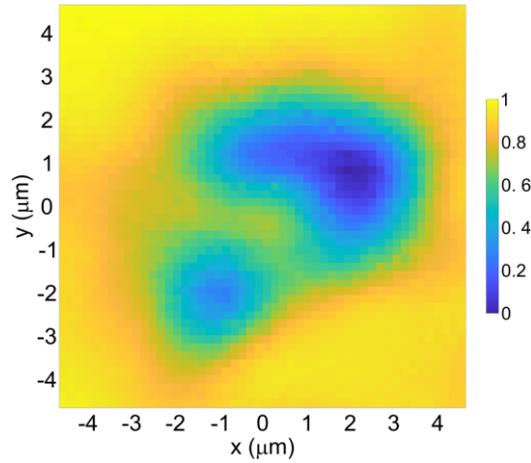

**Figure S11. Spatially dependent renormalized potential profile.**

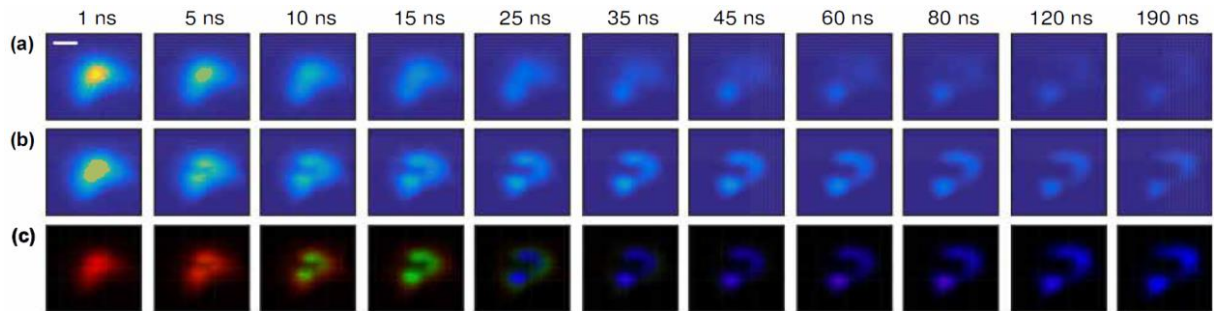

**Figure S12. Time-resolved spatial distribution of IX modes and their simulation (more data). (a)** Experimental result. The data is normalized to the maximum of the entire evolution. A 726 nm pulsed excitation (1 MHz repetition) at 3.4 µW average power is used. Each frame is the spatial distribution of the PL signal with an exposure time of 64 ps. The scale bar is 2.5 µm. **(b)** Simulation result. The simulated spatial distribution is obtained by summing up the contribution from all modes together before normalizing it to the global maximum of the evolution. **(c)** Simulated mode contribution. The relative magnitude of the three modes, which are colour coded as red, green, and blue, for modes 1, 2, and 3, respectively, normalized to the maximum intensity of each frame.

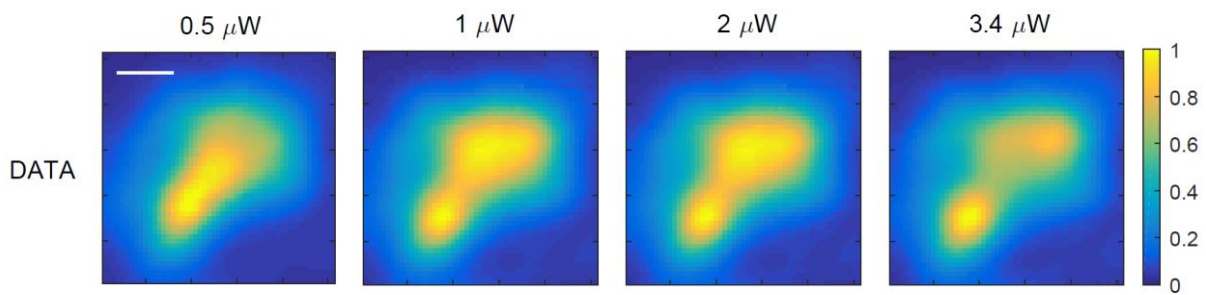

**Figure S13. Power-dependent spatial distribution of the PL emission at 50 mK.** Each map is taken by sweeping the collection path across the sample while keeping the excitation spot (of <1 μm diameter) fixed at the centre of the map. The sample is excited by a 726 nm pulsed laser (1 MHz repetition rate) with an average power labelled on the top. The scale bar is 2.5 μm. The data at each plot is normalized to the maximum PL count at that plot. Like in 4 K, IX spatial distribution resembles the high (low)-energy mode 1 (3) at low (high) excitation power. This result shows that, as our model suggested, this behavior is related to the nonlinear cascade term instead of the thermal contribution.

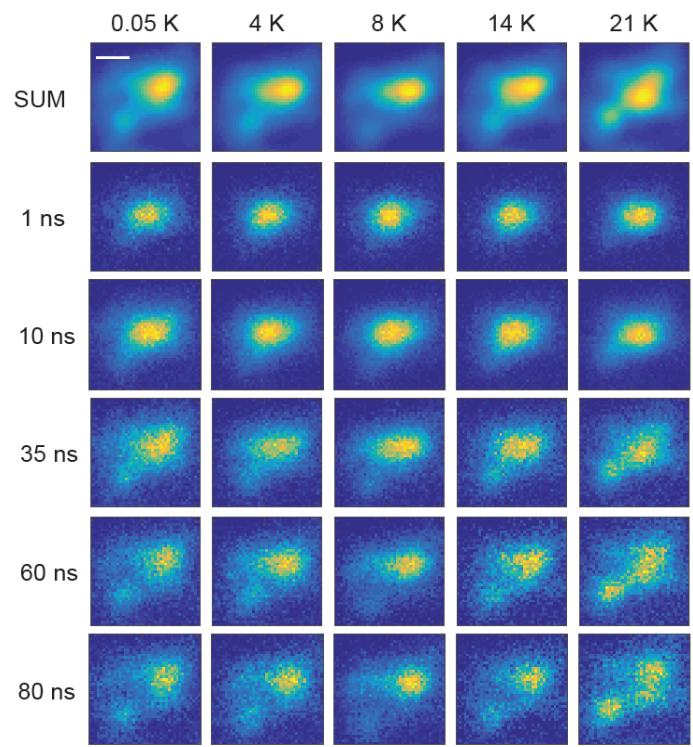

**Figure S14. Time-resolved spatial distribution of PL intensity at various temperatures.** The data is normalized to the maximum of the entire evolution. A 726 nm pulsed excitation (1 MHz repetition) at 3.4 μW average power is used. Each frame is the spatial distribution of the PL signal with an exposure time of 64 ps. The scale bar is 2.5 μm. The PL intensity spatial distribution is similar from 50 mK to 21 K, showing that this evolution is due to the cascade transition between these modes. Since this transition mainly depends on phonon emission instead of phonon absorption, it has a weak dependence on the temperature (within this temperature range).

| Parameter (unit) | Value at low power | Value at high power |
|---|---|---|
| $D$ (cm$^2$/s) | 0.06 | 0.06 |
| $\Gamma_1^{-1}$ (ns) | 945 | 964 |
| $\Gamma_2^{-1}$ (ns) | 1214 | 1056 |
| $\Gamma_3^{-1}$ (ns) | 8452 | 9142 |
| $r_{12}^{-1}$ (ns) | 17.6 | 21 |
| $r_{23}^{-1}$ (ns) | 74 | 90 |
| $r_{13}^{-1}$ (ns) | 384.4 | 397 |

**Table S1. Fitting parameter values.**

**References**

1    Tang, Y. *et al.* Simulation of Hubbard model physics in WSe$_2$/WS$_2$ moiré superlattices. *Nature* **579**, 353-358, doi:10.1038/s41586-020-2085-3 (2020).

2    Kulig, M. *et al.* Exciton diffusion and halo effects in monolayer semiconductors. *Phys. Rev. Lett.* **120**, 207401, doi:10.1103/PhysRevLett.120.207401 (2018).

3    Park, H. *et al.* Dipole ladders with large Hubbard interaction in a moiré exciton lattice. *Nat. Phys.*, doi:10.1038/s41567-023-02077-5 (2023).

4    Xiong, R. *et al.* Correlated insulator of excitons in WSe$_2$/WS$_2$ moiré superlattices. *Science* **380**, 860-864, doi:doi:10.1126/science.add5574 (2023).